\documentstyle[epsfig]{mn2e}

\title[Empirical distributions of $\lambda$]{Empirical distributions of 
galactic $\lambda$ spin parameters from the SDSS}

\author[X. Hernandez, Changbom Park, B. Cervantes-Sodi and Yun-Young Choi]
{X. Hernandez$^1$, Changbom Park$^2$, B. Cervantes-Sodi$^1$, and Yun-Young Choi$^2$\\
$^1$ Instituto de Astronom\'\i a,
Universidad Nacional Aut\'onoma de M\'exico
A. P. 70--264,  M\'exico 04510 D.F., M\'exico \\
$^2$ Korea Institute for Advanced Study, Dongdaemun-gu, Seoul 130-722, Korea\\
}
\date{\today}
\begin{document}
\maketitle
\begin{abstract}

Using simple dimensional arguments for both spiral and elliptical galaxies, we present
formulas to derive an estimate of the halo spin parameter $\lambda$ for any real galaxy,
in terms of common observational parameters. This allows a rough estimate of $\lambda$,
which we apply to a large volume limited sample of galaxies taken from the SDSS data base.
The large numbers involved (11,597) allow the derivation of reliable $\lambda$ distributions,
 as signal adds up significantly in spite of the errors in the inferences
for particular galaxies. We find that if the observed distribution of $\lambda$ 
is modeled with a log-normal function, as often done for this distribution in dark matter
halos that appear in cosmological simulations, 
we obtain parameters $\lambda_{0}=0.04 \pm 0.005$ and 
$\sigma_{\lambda}=0.51 \pm 0.05$, interestingly consistent with values derived from
simulations. For spirals, we find a good correlation between empirical values of $\lambda$
and visually assigned Hubble types, highlighting the potential of this physical parameter
as an objective classification tool.

\end{abstract}

\section{Introduction}
Perhaps one of the most useful tools for describing the nature of observed galactic populations is the
luminosity function. The relative numbers of galaxies of different luminosities offer a 
quantitative description of the result of the structure formation scenario of the universe, nowadays
available for inspection even as a function of redshift. The luminosity function, or mass
function of galaxies, once a modeling of star formation histories and dust is included to 
yield mass to light ratios, is indeed one of the principal constraints against which cosmological
simulations and structure formation theories in general are tested. Often star formation efficiencies
and histories are calibrated through a fitting of predicted halo mass functions and observed
luminosity functions e.g. Davis et al. (1985) and references thereof. 

Galaxies however, have many more properties than just their mass, most notably, 
galactic classification schemes center on sorting galaxies as ellipticals, lenticulars and spirals,
subdividing the latter again into types, typically along the Hubble sequence. In spite
of the many well known virtues of this classification scheme, its somewhat subjective,
qualitative and relative nature, has made it difficult to use in comparisons against cosmological
simulations, where the 'type' of a modeled galaxy is rather difficult to assess, in terms of
Hubble's classification scheme.

Inspired by varied theoretical studies which invariably identify the $\lambda$ spin parameter of
a host halo as the principal physical parameter in determining the morphological and visual
characteristics of a spiral galaxy, which are then subjectively integrated into the qualitative
assignment of 'type', in Hernandez \& Cervantes-Sodi (2006) two of us derived a simple estimate
of $\lambda$ for any observed spiral. There it was shown that the scalings of the derived  
$\lambda$ against various type determining properties such as colour, disk thickness and
bulge to disk ratio, are comparable to the corresponding scalings of these parameters against
Hubble type, using a large sample of nearby spirals.

It is interesting that a generic prediction of cosmological N-body simulations over the past 2 decades
has been the functional form and parameters which define the predicted distribution
of $\lambda$ for dark matter halos e.g. Bullock et al. (2001). Nevertheless, it has not been easy to test 
this prediction directly,
as $\lambda$ is not a straight forward observable feature of a real galaxy. 
Not only is a measurable $\lambda$ distribution relevant as a test of the general structure 
formation scheme, but also as a further way of independently constraining cosmological parameters, 
which to a certain extent alter its details e.g. Gardner (2001). 

Having the approximate 
estimates of Hernandez \& Cervantes-Sodi (2006) (henceforth HC06), here we set out to measure
empirical distributions of galactic $\lambda$ spin parameters. Firstly, given the approximate nature of 
these estimates, the only way of having signal adding up to something significant is through
the use of very large samples. Also, if we want a meaningful comparison against cosmological
models, use of a volume limited sample is crucial. These two constraints drive us inevitably to
the Sloan Digital Sky Survey (SDSS). Two of us have worked extensively with this database,
and recently constructed an interesting morphology segregation scheme of galaxies in the
SDSS in Park \& Choi (2005), henceforth PC05. Here we use a volume limited sample from
the SDSS having galaxies with redshifts in the interval $0.025 < z < 0.055$, corresponding
to distances of between $75h^{-1}$ Mpc and $162.57h^{-1}$ Mpc, assuming a WMAP cosmology of $\Omega_{M}=0.27$,
$\Omega_{\Lambda}=0.73$ and $h=0.71$, which we keep throughout. 

The following section includes a review of the derivation of $\lambda$ spin parameters for
spirals of HS06, and gives our results for the spirals in our SDSS sample. The correspondence
between $\lambda$ for spirals, galactic type through visual inspection and the colour and colour gradient
morphology segregation scheme of PC05 is also made in Section 2, where we also present empirical distributions
of galactic $\lambda$ spin parameters for spirals. These are well fitted by a log-normal function, 
and give parameters in accordance with recent cosmological simulations. In section 3 we extend the
dimensional estimate of $\lambda$ for ellipticals, and construct and analyze the empirical distributions of
$\lambda$ for the complete sample. Our results are summarized in 4.

\section{$\lambda$ distributions for spiral galaxies}   
We are concerned with the determination of the dimensionless angular momentum parameter of
a galactic halo:

\begin{equation}
\label{Lamdef}
\lambda = \frac{L \mid E \mid^{1/2}}{G M^{5/2}}
\end{equation}

where $E$, $M$ and $L$ are the total energy, mass and angular momentum of the configuration, respectively.
None of the quantities which appear in the above definition are accessible to direct observation.
In HC05 we derived a simple estimate of $\lambda$ for disk galaxies in terms of observational
parameters, which we summarize below.

We shall model only two galactic components, the first one a disk having a surface mass 
density $\Sigma(r)$ satisfying:

\begin{equation}
\label{Expprof}
\Sigma(r)=\Sigma_{0} e^{-r/R_{d}},
\end{equation} 

Where $r$ is a radial coordinate and $\Sigma_{0}$ and $R_{d}$ are two constants which are allowed to vary from
galaxy to galaxy. The total disk mass is now:

\begin{equation}
M_{d}=2 \pi \Sigma_{0} R_{d}^{2}.
\end{equation} 

The second component will be a fixed dark matter halo having an isothermal $\rho(r)$ density 
profile, and responsible for establishing a rigorously flat rotation curve $V_{d}$ throughout the entire galaxy,
an approximation sometimes used in simple galactic evolution models e.g. Naab \& Ostriker (2005).
Under a flat rotation curve the disk specific angular momentum from eq(\ref{Expprof}) 
will be $l_{d}=2 V_{d} R_{d}$, and the halo density will satisfy:

\begin{equation}
\label{RhoHalo}
\rho(r)={{1}\over{4 \pi G}}  \left( {{V_{d}}\over{r}} \right)^{2}, 
\end{equation}

having a halo mass profile $M(r)\propto r$. We define the disk mass fraction as
$F=M_{d}/M_{H}$, expected to be of order $1/10$ or smaller (e.g. 
Flores et al. 1993, Hernandez \& Gilmore 1998), we shall use global 
parameters for the entire system indistinctly from halo
parameters, consistent with having ignored disk self gravity in eq(\ref{RhoHalo}).

We must now express $\lambda$ in terms of structural galactic parameters readily
accessible to observations. First we assume that the total potential energy of the galaxy
is dominated by that of the halo, and that this is a virialized gravitational structure,
which allows to replace $E$ in equation(\ref{Lamdef}) for $W/2$, one half the gravitational
potential energy of the halo, given by:

\begin{equation}
\label{Epot}
W=-V_{d}^{2}M_{H}.
\end{equation}

Assuming that the specific angular momenta of disk and halo are equal, $l_{d} =l_{H}$
(as would be the case for an initially well mixed protogalactic state, and generally 
assumed in galactic formation models e.g. Fall \& Efstathiou 1980, Mo et al. 1998)
we can replace $L$ for $M_{H} l_{d}$. Introducing this last result
together with eq(\ref{Epot}) into eq(\ref{Lamdef}) yields:

\begin{equation}
\label{Lamhalo}
\lambda=\frac{2^{1/2} V_{d}^{2} R_{d}}{G M_{H}}.
\end{equation}

Finally, we can replace $M_{H}$ for $M_{d}/F$, and introduce a disk Tully-Fisher relation:

\begin{equation}
\label{TullyF}
M_{d}=A_{TF} V_{d}^{3.5}
\end{equation}

into eq(\ref{Lamhalo}) to yield:

\begin{equation}
\label{LamTully}
\lambda= \left( \frac {2^{1/2} F}{A_{TF}} \right) \frac{R_{d}}{G V_{d}^{3/2}}.
\end{equation}

The existence of a general baryonic Tully-Fisher relation between the total baryonic component and $V_{d}$ 
of the type used here, rather than a specific Tully-Fisher relation involving total magnitude in any particular band,
is supported by recent studies, in general in agreement with the 3.5 exponent we assume (e.g. Gurovich et al. 2004
or Kregel et al. 2005 who find 3.33 $\pm$ 0.37).
All that remains is to choose numeric values for $F$ and $A_{TF}$, to obtain an estimate of the
spin parameter of an observed galaxy in terms of structural parameters readily accessible to
observations, $R_{d}$ and $V_{d}$.

Taking the Milky Way as a representative example, we can use a total baryonic mass of
$1 \times 10^{11} M_{\odot}$ and $M_{H} = 2.5 \times 10^{12} M_{\odot}$ (where the estimate of the
baryonic mass includes all such components, disk, bulge, stellar halo etc., e.g. Wilkinson \& Evans 1999, 
Hernandez et al. 2001) as
suitable estimates to obtain $F=1/25$. For a rotation velocity
of $V_{d}=220 km/s$, the above numbers imply $A_{TF}=633 M_{\odot} (km/s)^{-3.5}$, this in turn
allows to express eq(\ref{LamTully}) as:

\begin{equation}
\label{LamObs}
\lambda=21.8 \frac{R_{d}/kpc}{(V_{d}/km s^{-1})^{3/2}},
\end{equation}

For the Galactic values used above, equation (\ref{LamObs})
yields $\lambda_{MW}=0.0234$, in excellent agreement with recent estimates of this parameter, e.g.
through a detailed modeling of the Milky Way within a CDM framework Hernandez et al. (2001) find
$\lambda_{MW}=0.02$.

First, we note that by construction, what we are estimating is not
strictly $\lambda$, but what has been defined as $\lambda'$, the equivalent $\lambda$ for a singular 
truncated isothermal halo. The relation between  $\lambda$ and $\lambda'$ is slightly a function of 
halo structure, for example, for NFW profiles with concentration of 10, one gets $\lambda'=0.9 \lambda$
e.g. Mo et al. (1998).
This slight difference is smaller than the error introduced through the dispersion in the Tully-Fisher relation
used, and will not be considered further, in any case, it is more rigorously $\lambda'$ and not $\lambda$
that we will be talking about. Also, it is clear that eq(\ref{LamObs}) is at best a first order approximation,
and not a precise evaluation of $\lambda$ for a real galaxy. 

However, in HC06 it was shown that this $\lambda$
parameter shows a one-to-one correlation, with small dispersion, when compared to the input
$\lambda$ of detailed galactic formation models from two distinct cosmological groups. Also,
the scalings seen against colour, disk thickness and bulge to disk ratios, for a sample of nearby
spirals from Courteau (1996, 1997), de Grijs (1998) and Kregel et al. (2002), 
similar to what is seen against Hubble type, highlights the use of this parameter 
as a physical classification scheme. This is not
surprising, as it is $\lambda$ as defined by eq.(\ref{Lamdef}) what has been repeatedly identified
in analytic and numerical studies of galactic formation as the principal determinant of galactic
type e.g. Fall \& Efstathiou (1980), Flores et al. (1993), Firmani et al. (1996), Dalcanton et al. (1997), 
Zhang \& Wyse (2000), Silk (2001), Kregel et al. (2005).

For example, the ratio of disk scale height to disk scale length, $h/R_{d}$, is 
one of the type defining characteristics 
of a galaxy which it is easy to show, will scale with $\lambda$. 
Starting from Toomre's stability criterion:

\begin{equation}
\label{Toomre}
Q(r)=\frac{\kappa(r) \sigma(r)}{\pi G \Sigma(r)},
\end{equation}

were $\kappa(r)$ and $\sigma(r)$ are the epicycle frequency and velocity dispersion at a given point in the
disk, assuming a thin disk, virial equilibrium 
in the vertical direction (Binney and Tremaine 1994) yields a relation between $h$ and $\Sigma$,

\begin{equation}
\label{VertEq} 
h= \frac {\sigma_{g}^{2}} { 2 \pi G \Sigma}. 
\end{equation}

We use this relation for 
$h$ to replace the gas velocity dispersion appearing in equation (\ref{Toomre}) for a combination of 
$h$ and the surface density. The dependence on $\Sigma$ is replaced by one on $M_{d}$
and $R_{d}$ through the disk profile. Replacing $\kappa(r)$ for $\sqrt{2} V_{d}/r$, and $V_{d}$
for $\lambda$, $R_{d}$ and $M_{H}$ through eq(\ref{Lamhalo}) to get a new expression for the 
Toomre's stability criterion, which evaluating radial dependences at $r=R_{d}$ yields,

\begin {equation}
Q^2= e 2^{5/2} \left( \frac{M_{H}}{M_{d}} \right) \left( \frac{h}{R_{d}} \right) \lambda
\end {equation}

With $F=1/25$, evaluating at $Q=1$, the stability threshold suggested by self-regulated star formation cycles, 
(e.g.  Dopita \& Ryder 1994, Koeppen et al. 1995 and Silk 2001) the ratio $h/R_{d}$ is obtained as:

\begin {equation}
\label{hRratio}
\frac{h}{R_{d}} =  \frac{1}{390 \lambda},
\end {equation}

a simplified version of the result of van der Kruit (1987).
For the Galactic value derived above of $\lambda_{MW}=0.0234$, equation (\ref{hRratio}) gives
$R_{d}=9 h$, not in conflict with parameters for the Milky Way.
For galaxies with large values of $\lambda$ we expect thin systems, while galaxies with small values of $\lambda$ 
will show thick disks, as the observed trend of ($h/R_{d}$) decreasing in going from early-type disks
to late type galaxies e.g. de Grijs (1998), Yoachim \& Dalcanton (2006).

Having a way of estimating $\lambda$, we now need to define the optimal sample to
use. As already mentioned, requiring as large as possible a volume limited sample
makes the SDSS an ideal database. Since most published cosmological distributions of 
$\lambda$ report data at $z=0$, we will start by taking a 
volume-limited, absolute magnitude limited sample with $M_{r}- 5$ $log$ $h \le -18.5$,
in a low redshift range, 
$0.025<z<0.055$, allowing an optimal comparison with reported cosmological studies. 
This sample contains 31,685 galaxies for which exponential disk scales, absolute magnitudes, 
velocity dispersions, de Vacouleurs radii and eccentricities have been determined (Choi et al. 2006). 
It is important to note that the above structural parameters were determined in red and IR
bands, in the spirit of obtaining global mass distribution parameters, it is clear that 
this is what is required for the measure of $R_{d}$ necessary in eq(\ref{LamObs}).

The first step is to identify the spiral galaxies in our sample. 
This is done using the colour, colour-gradient and concentration criteria
developed in PC05, where extensive testing and corroboration of the morphological segregation
criteria were preformed against large training sets of visually classified galaxies.
This procedure yields 21,184 spirals and 10,501 ellipticals.

Since we have no rotation curves for the spiral galaxies, as required by
equation(\ref{LamObs}), we must infer this through the absolute magnitudes of the observed
systems, and use of an appropriate Tully-Fisher relation. Hence, internal absorption in
edge-on disks is a problem we must avoid. We prune the original sample to leave only
spiral galaxies having axis ratios $>0.6$, this ensures only relatively face-on disks
remain, minimizing errors in the absolute magnitudes used (see Choi et al. 2006
for the choice of axis ratio cut). Also, we use the red band
Tully-Fisher relations of Barton et al. (2001) to assign rotation velocities to observed
galaxies. This relation shows a good fit to the data only in the range $-20>M_{R}>-22.5$,
so we must limit our galaxies to those falling in this range. The corresponding ranges we
impose on inferred rotation velocities are $80<V_{R}<430$ in km/s, well within the range
of applicability of the Tully-Fisher relation we are using. Still, the dispersion in this
Tully-Fisher, plus that in the baryonic Tully-Fisher used in deriving eq.(\ref{LamObs}),
leave us with a 25\% uncertainty in our individual estimates of $\lambda$ for spirals.

\begin{figure}
\psfig{file=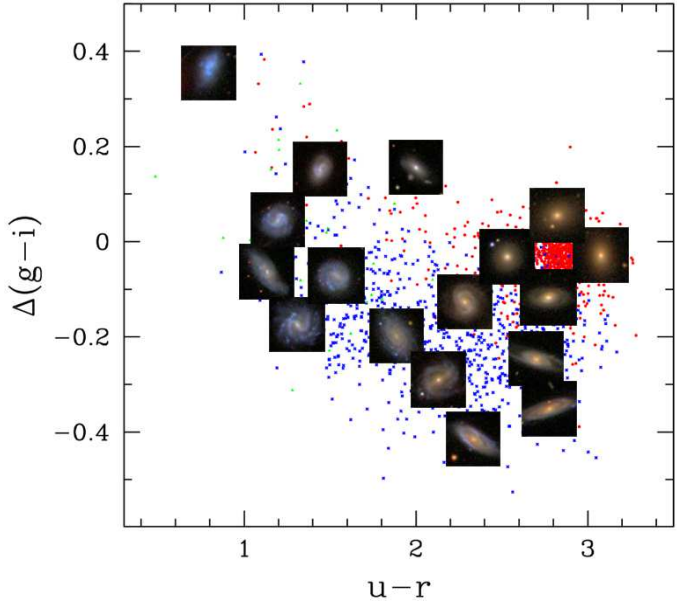,angle=0,width=9.0cm,height=8.0cm}
\caption{Some of the galaxies in our sample on a colour, colour-gradient plane,
with photos of galaxies representative of those found in each region, as visually 
classified by Park \& Choi (2005). Red dots are early type galaxies and others are late types.
} 
\label{fig1}
\end{figure}

After applying the two cuts described above, we are left with 0.366 of the original
21,184 spirals, 7,753 disk systems, still large enough to yield significant statistical information on the
distribution, and small enough to include only galaxies for which our estimates of $\lambda$
are most accurate. Use of eq.(\ref{LamObs}) for this sample yields a collection of 7,753 
values of $\lambda$ for a complete volume limited SDSS sample.

Having now a set of inferred, observational values of $\lambda$, we first explore the
correlations of $\lambda$ for spirals only, against the colour and colour-gradient
information found by PC05 to be an accurate indicator of galactic type. 
Figure (1), taken from PC05, shows some of the galaxies in our sample, with photos
of galaxies representative of those found in each region, as visually classified by PC05.
Notice the well defined cluster of ellipticals centered on $\Delta(g-i)=-0.04$, $u-r=2.82$.
Figure (2) shows all our spirals in a colour, colour-gradient plane, with the shading giving the average
values of $\lambda$ within each shaded square. We see that disks with high values of 
$\lambda$ are found in the lower left hand area, whilst disks having low values of $\lambda$
populate the right and upper regions. It is interesting that PC05 find precisely the same segregation pattern
for late and early spirals, respectively, reinforcing the results of HC06 that the quantitative
and objective $\lambda$ parameter constructed in eq.(\ref{LamObs}) is a good physical classification 
parameter for spirals, reproducing the broad trends of the classical subjective, qualitative
Hubble sequence.

Next we use our collection of values of $\lambda$ from our pruned sample of
face-on spirals to construct a histogram, shown in figure (3) by the broken curve.
The shape of this histogram suggests a log-normal distribution:

\begin{figure}
\psfig{file=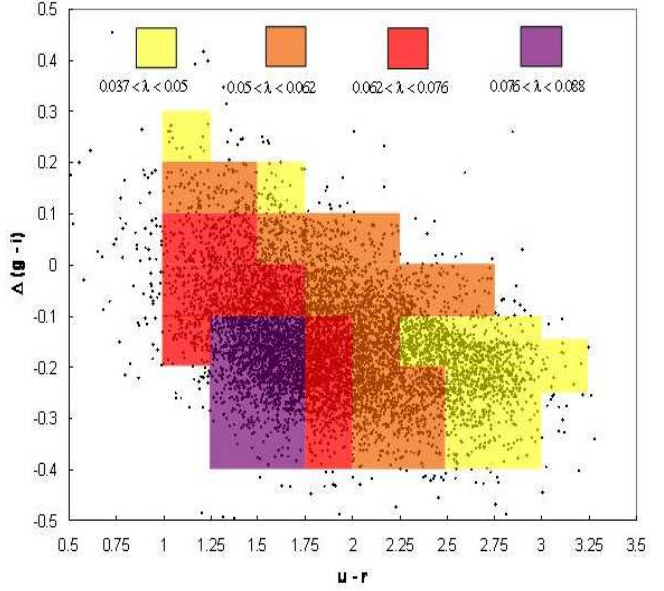,angle=0,width=9.0cm,height=8.3cm}
\caption{Spiral galaxies in our sample on a colour colour-gradient plane, the 
shading shows the average values of $\lambda$ in each shaded region, using eq.(\ref{LamObs}).
Regions of high and low values of $\lambda$ closely correspond to regions populated
by late type and early type spirals, respectively, as visually classified by Park \& Choi (2005).
} 
\label{fig1}
\end{figure}

\begin{equation}
\label{Plam}
P(\lambda_{0},\sigma_{\lambda};\lambda) d\lambda=
\frac{1}{\sigma_{\lambda}\sqrt{2\pi}}exp\left[-\frac{ln^{2}(\lambda/\lambda_{0})}
{2\sigma_{\lambda}^{2}} \right] \frac{d\lambda}{\lambda}
\end{equation}

The moments of the above distribution are analytical, which allows one to express:

\begin{equation}
M_{1}(\lambda_{0}, \sigma_{\lambda})=\frac{\int_{0}^{\infty} P(\lambda) \lambda d\lambda} 
{\int_{0}^{\infty} P(\lambda) d\lambda}=\lambda_{0} exp(\sigma_{\lambda}^{2}/2),
\end{equation}

\begin{equation}
M_{2}(\lambda_{0}, \sigma_{\lambda})=\frac{\int_{0}^{\infty} P(\lambda) \lambda^{2} d\lambda} 
{\int_{0}^{\infty} P(\lambda) d\lambda}=\lambda_{0}^{2} exp(2 \sigma_{\lambda}^{2}),
\end{equation}

from which one solves:

\begin{equation}
\lambda_{0}=\frac{M_{1}^{2}}{M_{2}^{1/2}},  
\sigma_{\lambda}=\left[ ln\left( \frac{M_{2}}{M_{1}^{2}}    \right) \right]^{1/2}.
\end{equation}

\begin{figure}
\psfig{file=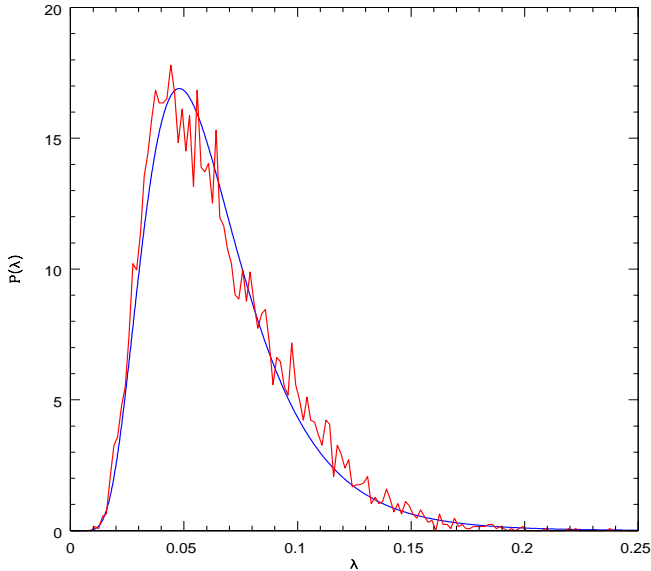,angle=0,width=9.0cm,height=8.3cm}
\caption{Distribution of values of $\lambda$ for 7,753 spirals using eq.(\ref{LamObs}), 
binned into 150 intervals, broken curve. Best log-normal fit to the data, having parameters 
$\lambda_{0}=0.0585$ and $\sigma_{\lambda}=0.446$, solid curve.
} 
\label{fig1}
\end{figure}

Equating the 1st and 2nd order moments of our empirical distribution to $M_{1}$
and $M_{2}$, we find the best fit log-normal curve to our distribution for
spiral galaxies. This is given by the smooth curve in figure (3), where the empirical
histogram has been rescaled to a unit integral. The parameters of the best fit
distribution are: $\lambda_{0}=0.0585$, $\sigma_{\lambda}=0.446$. Both the form of this
distribution and the particular values for the parameters we find are in good agreement with
results from cosmological simulations, e.g. Shaw et. al (2006).

From figure (3) it is apparent that the log-normal functional form is a good description of our inferred data, 
a more formal confirmation is provided by a K-S test. We construct the cumulative distribution functions for both
the log-normal $P(\lambda)$ and our data out to $\lambda=0.2$, shown in figure (4) by the solid and dotted 
curves, respectively. These yield a maximum difference $D=0.029$, and a significance given by:

\begin{equation}
Q_{KS}(X)=2 \sum_{j=1}^{\infty} (-1)^{j-1} exp(-2j^{2}X^{2})
\end{equation}

where $X=N^{1/2}D$, $N$ the total number of points in our sample, 7,746,
giving $X=2.55$ and a significance level of $Q_{KS}=0.000004$.
We hence see that our inferred distribution of galactic $\lambda$'s is consistent with having come from 
the best fit log-normal $P(\lambda)$, to a very high degree. We note the important precedent of Syer et. al (1999)
who using similar estimates for $\lambda$ for spiral galaxies, using a sample from Lauberts (1982), 
find a distribution of values of $\lambda$ well fitted by a log-normal function, with parameters 
$\lambda_{0}=0.05$ and $\sigma_{\lambda}=0.36$. The small differences with our inference might come
from their lack of a large volume limited unbiased sample, their lack of an 
attempt at modeling $\lambda$ for ellipticals, which further limited the scope of comparisons which 
could then be made against cosmological n-body models, or their use of a disk formation model
calibrated directly from cosmological n-body simulations.

From equation(\ref{LamTully}) we see that our inferred $\lambda$'s are proportional to the constants
$(F/A_{TF})$, since the validity and constancy of the T-F relation is well established, we can think of our
inferred $\lambda$'s as $\lambda(25F)$, where $F$ is the baryon fraction of a spiral system. We have shown that
under the simplest assumption of a constant $F=1/25$, a log-normal distribution results from the data, clearly,
any other constant $F$ will also yield a log-normal distribution, with the same $\sigma_{\lambda}$, and a 
mean of $25 F \lambda_{0}$. Also, notice that most cosmological simulations yield 
distributions of $\lambda$ being independent of mass 
(e.g. Bullock et al. 2001) which if true,
ensures our inferences will not be skewed by the velocity cuts applied to the spirals.

A baryon fraction being a strong function of $R_{d}$ or $V_{d}$ however,
could easily result in a best fit $P({\lambda})$ no longer looking like a log-normal distribution.
In the absence of any definitive handle on $F$ or its possible variations, we limit ourselves to
concluding that a constant $F$ implies a log-normal $P(\lambda)$ for the spirals from the SDSS we
studied, while varying $F(R_{d}, V_{d})$ could doubtlessly imply something else.

\begin{figure}
\psfig{file=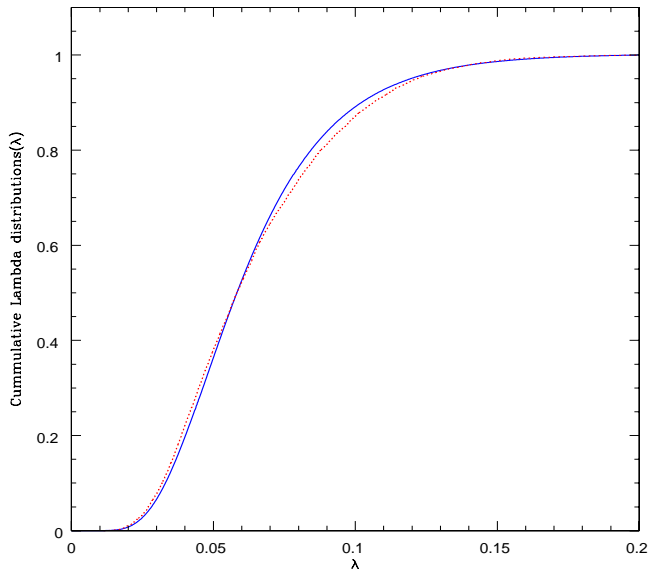,angle=0,width=9.0cm,height=8.3cm}
\caption{Cumulative distribution functions for both our spiral galaxies data and
the best fit log-normal $P(\lambda)$, dotted and solid curves, respectively. 
} 
\label{fig1}
\end{figure}

\section{$\lambda$ distributions for the complete sample}   
  
Before attempting an empirical total distribution of $\lambda$'s
we must derive an estimate for this parameter for an elliptical galaxy,
we propose a model equivalent to the one used for spirals,
including only two components; 
a baryonic matter distribution with a Hernquist density profile and 
a dark matter halo having a singular, truncated isothermal profile. 

In order to calculate the energy of the galaxy, we assume that the 
total potential is dominated by the energy of the virialized dark 
matter halo. In principle, we expect that
halos were elliptical galaxies are found are no different from those 
of disk galaxies (White \& Rees, 1978, Kashlinsky, 1982). In the case 
of a disk galaxy, the halo density profile has the form $4\pi G \rho(r) = \left(V_{c}/r\right)^{2}$, 
with potential energy $-V_{c}^{2}M_{H}$. The velocity
$V_{c}$ is the circular velocity of an element in centrifugal equilibrium with the halo. 
In an elliptical, nothing is actually moving at that velocity, which still can be
used to characterize the halo.  
The dependence on velocity can be changed to one on mass 
using again a baryonic Tully-Fisher relation (Gurovich et al, 2004):
$M_{b} = A_{TF}V_{c}^{3.5}$, thinking that the dynamics of a halo host of an elliptical galaxy are 
the same as those of disk galaxies. Finally, we define the baryonic mass 
fraction as $F = M_{b}/M_{H}$ to express the mass of the halo in terms of the 
baryonic mass. Validating the above assumptions, we note the interesting new
results of Treu et. al (2006) and Koopmans et. al (2006) who find trough dynamical studies of
velocity dispersion in ellipticals, and direct inferences on the halos of the studied
ellipticals through lensing measurements, 1) the presence of extensive dark halos 
and 2) that these halos have equivalent velocity dispersions equal to those of their
galaxies, the dynamical equivalent of a flat rotation curve for ellipticals. 

For the angular momentum, if we assume that the specific angular momenta 
of dark matter and baryons are equal, $l_{b}=l_{H}$, we can obtain the angular 
momentum of the entire configuration using only the directly observable 
component. It is important to remember that the baryonic component is 
susceptible of dissipating, while dark matter is 
not, this will affect the above assumption and our $\lambda$ estimates, being the value obtained a lower limit. 
In order to obtain the 
angular momentum of the baryonic distribution, we need to know the rotation 
velocity, information not always available due to 
technical observational difficulties and the fact that in many 
elliptical galaxies, the rotational velocity is at best of the same order as 
the velocity dispersion (Pinkney et al. 2003, Sarzi et al. 2006).
These forces us to determine the angular momentum of 
the system using some other observable parameter. In many works (e.g. Binney 1978, Franx et al. 1991) 
it is shown that the ellipticity of a galaxy is 
related to its rotation velocity, more precisely to its angular 
momentum. By dimensional analysis we expect the specific 
angular momentum of the galaxy be proportional to $(GM_{b}a)^{1/2}$ and 
to the eccentricity $e$ of the galaxy in the i-band, as shown by Gott \& Thuan (1976) 
and in Marchant \& Shapiro (1977). Introducing a proportionality 
constant  $C$ related to the details of the matter distribution, the specific angular 
momentum is given by:

\begin{equation}
\label{lEl}
l_{b} = C e \sqrt{GM_{b}a},
\end{equation} 
with $a$ the de Vacouleurs radius of the galaxy.
The constant C can be calibrated from cases where we know the actual
angular momentum of the system. For a small sample of 22 closely studied elliptical 
galaxies, from Halliday et al. (2001) and Pinkney 
et al. (2003) we calculated the actual angular momentum, for a Hernquist profile and using 
the reported values of rotational velocity. Comparing with the angular 
momentum given by eq.(\ref{lEl}), we obtain $C=0.114$.

For the galactic mass, using again the assumption that the gravitational 
structure is virialized, we compute the dynamical mass at $R_{50}$, the half light 
radius, using the expression 

\begin{equation}
M_{dyn}=\frac{(1.65\sigma)^{2}R_{50}}{G}
\end{equation}

from Padmanabhan et al. (2004), where $\sigma$ is the velocity dispersion. For 
elliptical galaxies it is well known that the presence of dark matter in 
the inner part of the galaxy is negligible so the total baryonic mass 
will be simply $2 M_{dyn}$.

Now we have the tools to estimate the energy, the angular momentum and the mass
of an elliptical galaxy. If we introduce this information into equation 1, we obtain:

\begin{equation}
\label{LamEl}
\lambda =\frac{0.051 e (a/kpc)^{2/7}}{(\sigma/km s^{-1})^{3/7}}.
\end{equation} 

Where we used values for $A_{TF}$ and $F$ as calibrated in HC06. 
The above is the equivalent expression of eq.(\ref{LamObs}) for elliptical galaxies,
and has errors of 30\%, due to the dispersion in the observational relations used. 

\begin{figure}
\psfig{file=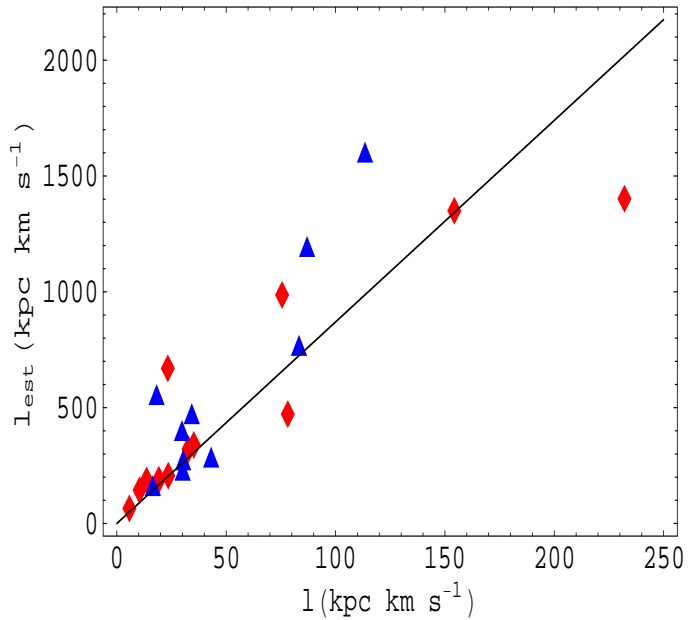,angle=0,width=9.0cm,height=8.3cm}
\caption{Specific angular momentum for elliptical galaxies from Pinkney et. al (2003) (triangles) and 
Halliday et. al (2001) (rhombi),
determined directly from their measurements of rotation velocity and deprojected light profiles, x axis.
The y axis shows values for the same quantity, but estimated through equation (\ref{lEl}) using $C=1$, the
solid line gives the best fit linear relation. We then estimate $C$ by forcing the best fit linear relation
to coincide with the identity.} 
\end{figure}

We must take into
account that the values for eccentricity $e$ and radius $a$, required
must be the intrinsic parameters of the system, not the
observed ones, as these will be systematically smaller due to projection effects.
Also in many ellipticals, shape is the result not exclusively of rotational support, but
also of dynamical pressure support (i.e. anisotropic velocity dispersion), 
with varying degrees of relative importance. Hence, unknown
dissipational and projection effects make eq.(\ref{LamEl}) a lower limit, whilst 
the participation of pressure support make it an upper one. To what extent the above compensate
each other is unclear, we shall therefore assume eq.(\ref{LamEl}) must be multiplied by 
a factor which remains a free parameter of our method, $X_{E}$.

Since ellipticals will generally have much lower values of
$\lambda$ than spirals, the relative mix of spirals and ellipticals will be an important factor in
determining our final estimated distribution of galactic $\lambda$'s. 
We must hence leave this relative fraction equal to that of the original
sample, before the velocity and projection cuts were applied to the spirals, 
we further remove randomly selected ellipticals to reduce the total number of galaxies to
0.366 of the original number, leaving only 3,843 ellipticals. Our final sample is composed of a
total of 11,597 galaxies, large enough to yield significant statistical information on the
distribution.

Using eq.(\ref{LamEl}) we estimate $\lambda$ for the ellipticals in the sample, after which
we can add the two distributions to obtain the total empirical $P(\lambda)$. As discussed
previously, for ellipticals, a degree of uncertainty remains, in connection to the (varying) degree of
pressure support, the degree of dissipation of angular momentum in the baryonic component, and
projection effects, which to 0th order we have incorporated as an unknown multiplicative factor
$X_{E}$. For $X_{E}=1$ the result of adding the distribution of $\lambda$ parameters for
spirals and ellipticals is a distinctly double peaked $P(\lambda)$, with the ellipticals coming in
as a gaussian distribution centered
at $\lambda=0.01$. In the interest of comparing with simulations, we shall take as a working
hypothesis that the full $P(\lambda)$ should also be a log-normal distribution. We then 
adjust $X_{E}$ so as to obtain a log-normal total distribution, with the restriction that the 
values of $\lambda$ for ellipticals should remain lower than those of spirals. The resulting histogram,
requiring $X_{E}=2.3$ is shown by the broken curve in figure(6), where the smooth solid curve gives
a best fit log-normal curve to the full distribution, having parameters $\lambda_{0}=0.046$
and $\sigma_{\lambda}=0.535$. 

Given that we have used empirical scaling relations, which show a measure of dispersion, 
we now take the pessimistic view that such dispersions are intrinsic to
the data (pessimistic in the sense that a measure of this dispersion is undoubtedly observational
noise), and hence, propagate as errors into our estimates of $\lambda$. We now estimate what is the
best underlying log-normal distribution, assuming that the histograms we obtained are only
a degraded version of an intrinsic log-normal distribution.
We determine the best intrinsic log-normal function through a full
maximum likelihood analysis of the data. This is done for purposes of comparison against
predicted $\lambda$ distributions, we fix the functional form of the distribution, and set out
only to evaluate the best fit parameters $\lambda_{0}$ and $\sigma_{\lambda}$ from eq.(\ref{Plam}).

We construct the likelihood function as the probability that an inferred set of $n$
empirical values of $\lambda$, $\{\lambda_{i}(\lambda)\}^{n}$, might arise from a given 
model, ($\lambda_{0}, \sigma_{\lambda}$), as:

\begin{equation}
\label{likelyhood}
{\cal {L}} \left( \lambda_{0}, \sigma_{\lambda};\{\lambda_{i}(\lambda)\}^{n} \right)=
\sum_{i=0}^{n} ln \int_{0}^{\infty} P(\lambda_{0}, \sigma_{\lambda};\lambda) \times 
\lambda_{i}(\lambda) d\lambda
\end{equation}

where:

\begin{equation}
\lambda_{i}(\lambda)=\frac{F_{i}}{\sigma_{i}\sqrt{2 \pi}} exp\left[-\frac{(\lambda-\lambda_{0i})^{2}}{2\sigma_{i}^{2}} 
\right],
\end{equation}

$\lambda_{0i}$ is the center of a density distribution, the nominal inferred $\lambda$ for the ith galaxy,
$\sigma_{i}$ the assigned error, and $F_{i}$ a normalization factor taking into account that no matter what the
error, the domain of $\lambda_{i}$ must always be the positive values of $\lambda$. Therefore, we normalize 
$\lambda_{i}(\lambda)$ as a truncated Gaussian having unit integral, giving $F_{i}=2/[{1+erf\left(\lambda_{i0} 
/\sqrt{2} \sigma_{i}  \right)}]$.

In the limit of the error in our inferred $\lambda$ tending to zero, $F_{i}$ tends to 1, and $\lambda_{i}(\lambda)$
tends to a Dirac delta function, reducing the integral in equation(\ref{likelyhood}) to an evaluation of
$P(\lambda)$ at the inferred value. The advantage of a full likelihood formulation is that parameter inference
can be preformed in a way which naturally incorporates the errors in the data sample, without the need of binning
the data, a process which intrinsically reduces the information content of the sample.

Equation(\ref{likelyhood}) is then evaluated over a fine grid of values of  
($\lambda_{0}$, $\sigma_{\lambda}$), and the point where the maximum is found selected
as the best fit model. For the 7,753 spirals the result is ($\lambda_{0}=0.0517$, $\sigma_{\lambda}=0.362$),
with very small confidence intervals of $\pm 0.0003$ and $\pm 0.004$, respectively. Although
our errors in individual $\lambda$ estimates are of 25\% of $\lambda$, it is thanks to having a sample 
running into the thousands that we can retrieve details of the $\lambda$ distribution with accuracy.

Finally, we repeat the evaluation of eq.(\ref{likelyhood}), but using this time the complete sample,
of 11,597 spirals plus ellipticals. Here $\lambda$ for the ellipticals is taken as $2.3 \times$
the estimate of eq.(\ref{LamEl}). 
This factor can not be much larger than what we are using, since then
ellipticals would start overlapping significantly with spirals in their $\lambda$ distributions, as mentioned
previously, if this
factor is much reduced, the total $\lambda$ distribution becomes double peaked, with ellipticals
appearing as a distinct population at very low $\lambda$, which would be hard to explain.

The results this time are $\lambda_{0}=0.0394$ and $\sigma_{\lambda}=0.509$.
A comparison of the maximum likelihood model and the collection of inferred $\lambda$s is given in
fig.(6), which is analogous to fig.(3), but includes the complete sample, binned into 150 discrete intervals.
A direct fit of eq.(\ref{Plam}) to the full data gives $\lambda_{0}=0.046$, $\sigma_{\lambda}=0.535$.

The formal errors in the method are again of the order of what was found for the spirals, but this time
we are dominated by the unknown correction factor between our estimate and the actual $\lambda$ for
ellipticals. 
This uncertainty, although bounded, dominates our final error estimates and yields
$\pm 0.005$ in $\lambda_{0}$ and $\pm 0.05$ in $\sigma_{\lambda}$. Another possible source of error
in our estimates would be the existence of a large population of low surface brightness galaxies
of high $\lambda$,
which would lead to larger values of both $\lambda_{0}$ and $\sigma_{\lambda}$.

\begin{figure}
\psfig{file=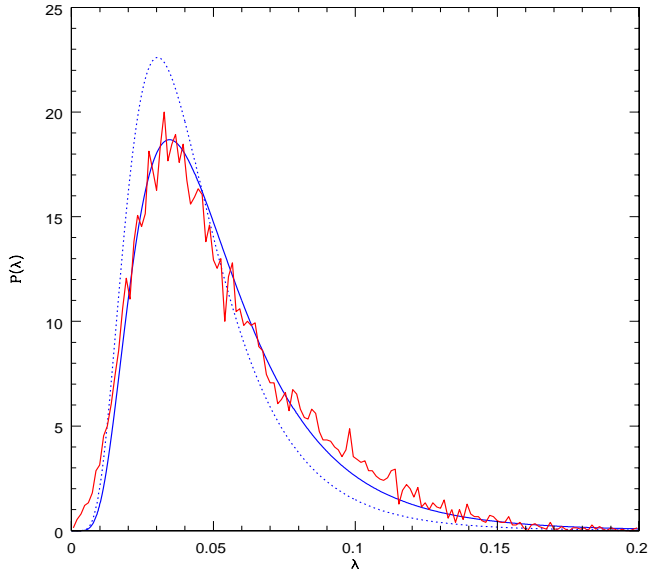,angle=0,width=9.0cm,height=8.3cm}
\caption{Distribution of values of $\lambda$ for 11,597 spirals plus
ellipticals using eq.(\ref{LamObs}) and eq.(\ref{LamEl}), binned into 150 intervals, broken curve.
Best log-normal fit to the data, having parameters 
$\lambda_{0}=0.046$ and $\sigma_{\lambda}=0.535$, smooth solid curve. Maximum likelihood
underlying log-normal distribution, assuming a 30\% error in our inferred $\lambda$'s, with parameters
$\lambda_{0}=0.0394$ and $\sigma_{\lambda}=0.509$, dotted curve.
} 
\label{fig1}
\end{figure}

The above results qualitatively agree with
generic predictions of structure formation models regarding the predicted values 
for $\lambda_{0}$ and $\sigma_{\lambda}$,
e.g. Shaw et al. (2006) review recent results from the literature giving values in the range
$0.03 < \lambda_{0}< 0.05$ and $0.48< \sigma_{\lambda}< 0.64$. The availability of empirical measurements
of these parameters could serve as a further independent guideline for models of structure formation.

We stress that for the full sample, the log-normal nature of both the inferred and the underlying
$\lambda$ distributions is an assumption which we can not confirm, taken only to permit a comparison with models.
Under this assumption, the parameters of the distribution can be derived from the data.
An inference of the actual functional form of the total distribution would require
more solid constraints on the hypothesis introduced in the derivation of the $\lambda$'s
for ellipticals. However, we do note that any similar scheme modeling ellipticals as residing in
halos equivalent to those of spirals, and having significantly lower angular momentum than spirals,
will also yield a $P(\lambda)$ not very different from what is shown in figure (6).

\section{conclusions}
We apply simple estimates of the $\lambda$ spin parameter to a large volume limited
sample from the SDSS. The sample is split into ellipticals and spirals using colour,
concentration and colour gradients. 

We find that for spiral galaxies, the average value of the inferred
$\lambda$ correlates well with standard type, as determined by visual inspection. 
This highlights the potential of $\lambda$ as an objective and quantitative
classification tool for spiral galaxies.

For the spiral galaxies sample we obtain a distribution which is statistically
consistent with a log-normal distribution, as what is commonly found in cosmological
n-body simulations, we find parameters $\lambda_{0}=0.0585$, $\sigma_{\lambda}=0.446$.

Ellipticals have, as expected, average values of $\lambda$ an order of magnitude lower than spirals.
The details of their $\lambda$ distribution are harder to quantify, but average values of
around $0.005$ are derived.

If the distribution of $\lambda$ parameters for the full sample is
fitted by log-normal distribution, of the type found to reproduce the corresponding distribution
of modeled halos arising in cosmological simulations, the parameters we find are:
$\lambda_{0}=0.0394 \pm 0.005$ and $\sigma_{\lambda}=0.509 \pm 0.05$, derived this time from a sample
of real galaxies.

\section{ACKNOWLEDGMENTS}
The work of X. Hernandez was partly supported by DGAPA-UNAM grant No IN117803-3 and CONACYT grants 
42809/A-1 and 42748. CBP is supported by the Korea Science and Engineering Foundation (KOSEF) through the 
Astrophysical Research Center for the Structure and Evolution of Cosmos (ARCSEC). 
The work of B. Cervantes-Sodi is supported by a CONACYT scholarship.

\end{document}